\documentclass[12pt]{iopart}
\usepackage[]{color}
\begin{document}

\title{Grounding Bohmian Mechanics in  Weak Values and Bayesianism}
\author{H. M. Wiseman}
\address{Centre for Quantum Dynamics, Griffith
University,  Brisbane,
Queensland 4111 Australia. }

\newcommand{\beq}{\begin{equation}}
\newcommand{\eeq}{\end{equation}}
\newcommand{\bqa}{\begin{eqnarray}}
\newcommand{\eqa}{\end{eqnarray}}
\newcommand{\nn}{\nonumber}
\newcommand{\nl}[1]{\nn \\ && {#1}\,}
\newcommand{\erf}[1]{Eq.~(\ref{#1})}
\newcommand{\erfs}[2]{Eqs.~(\ref{#1})--(\ref{#2})}
\newcommand{\dg}{^\dagger}
\newcommand{\rt}[1]{\sqrt{#1}\,}
\newcommand{\smallfrac}[2]{\mbox{$\frac{#1}{#2}$}}
\newcommand{\half}{\smallfrac{1}{2}}
\newcommand{\bra}[1]{\langle{#1}|}
\newcommand{\ket}[1]{|{#1}\rangle}
\newcommand{\ip}[2]{\langle{#1}|{#2}\rangle}
\newcommand{\sch}{Schr\"odinger}
\newcommand{\hei}{Heisenberg}
\newcommand{\ito}{It\^o }
\newcommand{\str}{Stratonovich }
\newcommand{\dbd}[1]{\frac{\partial}{\partial {#1}}}
\newcommand{\sq}[1]{\left[ {#1} \right]}
\newcommand{\cu}[1]{\left\{ {#1} \right\}}
\newcommand{\ro}[1]{\left( {#1} \right)}
\newcommand{\an}[1]{\left\langle{#1}\right\rangle}
\newcommand{\st}[1]{\left|{#1}\right|}
\newcommand{\implies}{\Longrightarrow}
\newcommand{\del}{\nabla}
\newcommand{\du}{\partial}
\newcommand{\singlecol}{\end{multicols}
     \vspace{-0.5cm}\noindent\rule{0.5\textwidth}{0.4pt}\rule{0.4pt}
     {\baselineskip}\widetext }
\newcommand{\doublecol}{\noindent\hspace{0.5\textwidth}
     \rule{0.4pt}{\baselineskip}\rule[\baselineskip]
{0.5\textwidth}{0.4pt}\vspace{-0.5cm}\begin{multicols}{2}\noindent}

\newcommand{\ea}{{\em et al.}}
\newcommand{\ps}[1]{\hspace{-5ex}{\phantom{\an{X_{w}}}}_{#1}}

\definecolor{nblue}{rgb}{0.2,0.2,0.7}
\definecolor{nblack}{rgb}{0,0,0}

\newcommand{\blu}{\color{nblack}}
\newcommand{\blk}{\color{nblack}}

\begin{abstract}
Bohmian mechanics (BM) is a popular interpretation of quantum mechanics in which particles have real positions. The velocity of a point ${\bf x}$ in configuration space is defined as the standard probability current ${\bf j}({\bf x})$ divided by the probability density $P({\bf x})$. 
However, this ``standard'' ${\bf j}$ is in fact only one of infinitely many that transform correctly and satisfy $\dot{P}+\del\cdot{\bf j}=0$. In this article I show that  a particular ${\bf j}$ is singled out if one requires that ${\bf j}$ be determined {\em experimentally} as a {\em weak value}, using a technique that would make sense to a physicist with no knowledge of quantum mechanics. This ``naively observable'' ${\bf j}$ seems the most natural way to define ${\bf j}$ 
operationally. Moreover, I show that this operationally defined
 ${\bf j}$ equals the standard ${\bf j}$, so, assuming $\dot{\bf x}={\bf j}/P$ one obtains  the dynamics of BM. It follows that the possible Bohmian paths are naively observable from a large enough ensemble. 
Furthermore, this \blu justification for the Bohmian law of motion \blk singles out ${\bf x}$ as the hidden variable, because (for example) the analogously defined  momentum current is in general incompatible with the evolution of the momentum distribution. Finally I discuss how, in this setting, the usual quantum probabilities can be  motivated  from a Bayesian standpoint, via the principle of indifference.
\end{abstract}


\maketitle

\section{Introduction} \label{sec:intro}

Hidden Variable (HV) interpretations of 
 quantum mechanics (QM)  are motivated by the desire to give a unified, realistic 
 description of the physical universe \footnote{ This is in contrast to the 
 Orthodox or Copenhagen interpretation, which introduces an ill-defined cut 
  between classical and quantum worlds, ruling out such a unified description. Originally, some, such as Einstein (see Refs.~\cite{Bel76a,Wis06a}) hoped that HVs would 
also solve another problem of the Copenhagen interpretation, that it it violates local causality. 
However Bell's theorem \cite{Bel6476,BellCollection} showed that this problem is inescapable for any interpretation of QM that admits the reality of measurement outcomes for space-like separated observers.} 
The earliest and most popular example of such a HV interpretation is that of de Broglie \cite{deB56}, further developed by Bohm \cite{Boh52}.  In de Broglie's formulation, the HVs were defined to be the position of particles.  Bohm rediscovered and refined this theory, and extended it by including the values of gauge fields at all points in space as HVs.  I will call this more general interpretation Bohmian mechanics (BM). Through the work of many people \cite{BohHil93,Hol93,CusFinGol96,Dur01} it has been established that BM gives a complete, and consistent picture of  reality in a quantum world.  It derives classical behaviour at the macroscopic level from microscopic dynamics,  and it reproduces all of the  predictions of Orthodox QM in the laboratory. As with every {\em interpretation} of QM it 
does not make any new predictions, but unlike some it is open to modifications that could lead to a deeper theory.

If BM is so successful,  why has it not been more widely adopted by physicists who seek a realistic description of nature?  Leaving aside socio-historical reasons \cite{Cus94,Wis06a}, there remains one good scientific reason: there are infinitely many other HV interpretations different from BM but achieving the same ends.  It was Bell who first devised a formalism that was not restricted to just position and field values \cite{Bel84}. Bell applied his dynamical model to fermion number, but his model allows a HV interpretation to be constructed from any variables that are sufficient to ascribe a definite state of experience to  a macroscopic observer \cite{Sud86}.

Bell's general theory has been related to the modal interpretation of QM \cite{Bub97}, and generalized further (See Ref.~\cite{GamWis04} and references therein). Not only can different HVs be chosen, but for a given choice of HVs, infinitely many different {\em dynamics} are possible. Part of this arbitrariness is related to an arbitrariness in the definition of the {\em probability current}. For example, for   non-relativistic QM the probability current for a 
single-particle of mass $m$  is conventionally defined as 
\beq \label{madelung}
{\bf j} ({\bf x};t) = ({\hbar}/{m}){\rm Im}\, \ip{\psi(t)}{{\bf x}} \mathbf{\nabla} \ip{{\bf x}}{\psi(t)}
\eeq
However, there are infinitely many other expressions for ${\bf j}$ that also obey the conservation equation 
\beq \label{conservation1}
\du P({\bf x};t)/\du t + \nabla\cdot {\bf j}({\bf x};t) = 0 
\eeq
for $P({\bf x};t) = \ip{\psi(t)}{{\bf x}} \ip{{\bf x}}{\psi(t)}$ while still satisfying ``all possible physically meaningful requirements one can put forward for them'' \cite{DeoGhi98}. 

It could be argued that the lack of a  unique HV interpretation is not a problem: since any number of them would solve the problems with the Orthodox interpretation, one could just accept that these solutions exist and go on using Orthodox interpretation in practice.  Against this position there are a number of points to consider. First,  physicists who believe that a HV interpretation is the best solution to these problems would presumably want to teach this in introductory quantum courses. For such pedagogy, having a unique theory to present would be a great advantage. Second, a HV model that was preferred over all others {\em on physical grounds} could aid our intuition into quantum phenomena. Third, such a HV interpretation might even provide hints towards a theory deeper than QM.

In this article I address the question of how one can justify preferring one HV interpretation  over all others on physical grounds. In Sec.~2 I show, in the context of non-relativistic QM (including fields), that a unique expression for the probability current ${\bf j}$ can be defined using the concept of {\em weak values}  \cite{AhaAlbVai88,RitStoHul91,Ste95,RohAha02,Brun03,Sol04,Gar04,Pry05}. Moreover, this means that the probability current (so defined) can be  {\em naively observed}. By this phrase I mean that  it can be measured experimentally using a technique that would make sense to a physicist with essentially no knowledge of QM  \cite{Gar04}.   Next, I show that this expression coincides with the conventional expression for ${\bf j}$. From this it requires only the assumption of determinism to \blu obtain \blk the dynamics of BM, as discussed in Sec.~3.   In addition, I will show in Sec.~4 how this route to BM singles out the configuration (rather than the momentum, for example) as the HV.  In Sec.~5 I discuss the nature of probability in BM, and in particular show that adopting the subjective view of probability, and applying the principle of indifference,  may  rid BM of its remaining arbitrariness. Sec.~6 relates the work in Secs.~2 and 4 to work by others, and Sec.~7 concludes.

\section{Weak Values and the Probability Current} \label{Sec:WVPC}
The probability current (\ref{madelung}) was introduced by Madelung in 1926 \cite{Mad26}, in the immediate aftermath of \sch's discoveries. Despite the lack of uniqueness discussed above \cite{DeoGhi98}, it remains the standard or text-book expression for ${\bf j}$. It is also used to  define the particle velocity in BM:
\beq \label{BMvelocity}
\dot{\bf x} = {\bf v}({\bf x};t) \equiv {\bf j}({\bf x};t)/P({\bf x};t).
\eeq 
 It has been argued \cite{DurGolZan96} that \erf{madelung} is the simplest expression satisfying certain physical requirements (such as Galilean invariance), but, as emphasized in Ref.~\cite{Dic98}, simplicity is not a property that can be rigorously defined. 

It might seem surprising that such a basic concept (in scattering theory for example), should rest on such a shaky foundation; one would think that ${\bf j}$ would be defined operationally. The central difficulty is of course that the probability current relates to the motion of a particle at a point, whereas the uncertainty relation forbids the simultaneous determination of position and momentum. However, this stricture is not as tight as it might seem. In particular, the uncertainty relation does not forbid sequential measurements of non-commuting quantities. If the first measurement were  absolutely precise, then there would be no point making the second. For example, a precise or {\em strong} measurement  of position would destroy all momentum information, so that the result of a subsequent position measurement would yield no information about the initial momentum.  However, QM allows for measurements that are not absolutely precise. In fact, in the opposite limit, of a {\em weak} measurement \cite{AhaAlbVai88}, the disturbance to the system state vector can be small. The price to be paid for such a lack of disturbance is in the quality of the measurement result: a  weak measurement result contains a large amount of noise. Thus a single result means essentially nothing; to obtain a good average for the measurement, the ensemble size must be large.

Consider a weak measurement of some observable $a$. 
A \emph{weak value}, denoted $\left\langle {\hat a_{\mathrm{w}}}\right\rangle_{\ket{\psi}}$ is the \emph{mean} value {from such weak measurements} on an ensemble of systems, each prepared in state $\ket{\psi}$. 
So far, the weak value is no different from the strong value, the 
ensemble mean value of strong or precise measurements. However, the weak value 
 differs from the strong value if one calculates the mean from a subensemble 
obtained by post-selecting only those results for which a later strong
measurement reveals the system to be in state $|{\phi }\rangle $. It is convenient to denote such a  weak value by $\ps{\bra{\phi}\!}\an{\hat a_{\mathrm{w}}}_{\ket{\psi}}$. Allowing unitary evolution described by $\hat U$ between the measurements, one finds \cite{AhaAlbVai88} 
\begin{equation}
\ps{\bra{\phi}\hat U}\an{\hat a_{\mathrm{w}}}{}_{\ket{\psi}\phantom{\hat U}\!\!\!\!}=\mathrm{Re}\frac{\langle {\phi 
}|\hat{U}\hat{a}|{\psi 
}\rangle }{\langle {\phi }|\hat{U}|{\psi }\rangle }.  \label{weakvalgen}
\end{equation}
For a derivation using the language of generalized measurements, 
see Ref.~\cite{Gar04}.
 Weak values have 
 been used to analyse a great variety of quantum phenomena \cite{RitStoHul91,Ste95,RohAha02,Brun03,Sol04,Gar04,Pry05}.

I now apply the concept of weak values to the probability current.\footnote{A completely unrelated derivation of trajectories for quantum systems based on weak values was given by Wang \cite{Wan97}. This is not a HV interpretation, but rather a formal relation between the \sch\ equation and stochastic trajectories for a {\em complex-valued} ``position''. } QM forbids us from determining  the velocity of a particle given that we know where the particle is. However, it does not forbid us from determining the mean value of that velocity, as a weak value. This can be done by making a weak measurement of velocity, followed immediately by a strong measurement of position. Since velocity is defined by the rate of change of position, in fact we want simply to make a weak measurement of initial position, then a strong measurement of position a short time $\tau$ later.
We then have the following {\em operational definition} for the velocity for a particle at position ${\bf x}$:
\beq
{\bf v}({\bf x};t) \equiv \lim_{\tau\to 0} \,\tau^{-1}\, {\rm E}[{\bf x}_{\rm strong}(t+\tau) - {\bf x}_{\rm weak}(t)|{\bf x}_{\rm strong}(t+\tau) = {\bf x}]. \label{opdefvel}
\eeq
Here E$[a|F]$ denotes the average of $a$ over the (post-selected) ensemble where $F$ is true.

 Consider a naive experimentalist, with no knowledge of quantum mechanics beyond the following basic facts about experiments at the microscopic scale:  (i) no matter how carefully a preparation procedure is repeated, the measured properties of the particle will vary in different runs of the experiment;  (ii) if a given property of the particle is measured strongly (arbitrarily accurately) then in general this will drastically alter the future distribution of measurement results;  (iii) if an arbitrarily weak measurement is used instead, the future distribution can remain essentially unaltered. For such an experimentalist, \erf{opdefvel} would be the only sensible way to measure the velocity of a particle at position ${\bf x}$ \footnote{ Strictly, the average velocity of the particle at this position; see Sec.~3. }.   Thus I will call the velocity \ref{opdefvel} is the {\em naively observable} velocity, and contend that it is the most natural operational definition of velocity. 

Let us now evaluate the naively observable velocity for a scalar particle for simplicity. Using the notation of \erf{weakvalgen}, Eq.~(\ref{opdefvel}) is related to a weak value as follows: 
\beq
{\bf v}({\bf x};t) = \lim_{\tau\to 0} \,\tau^{-1}\left[  {\bf x} - \ps{\bra{\bf x}\hat U(\tau)}\an{\hat{\bf x}_{\rm w}}_{\ket{\psi(t)}\phantom{\hat U}\!\!\!\!} \right].
\eeq
Here $\hat{\bf x}$ is the operator for position, $\ket{\bf x}$ is a position eigenstate, and $\ket{\psi(t)}$ is the state for that particle, and $\hat U(\tau) = \exp(-i\hat H\tau/\hbar)$,  where $\hat H$ is the Hamiltonian operator. Note that if (as in BM) the HV is the position ${\bf x}$, then the result of the strong measurement {\em is} the real particle position ${\bf x}$. From  \erf{weakvalgen}, a little manipulation reveals that this expression for the velocity evaluates to the real part of ${\bra{\bf x}i[\hat{H},\hat{\bf x}]\ket{\psi(t)}}/{\hbar \ip{\bf x}{\psi(t)}}$.  For ease of later generalization, this  is best stated as
\beq 
{\bf v}({\bf x};t) = {\rm Re}\frac{\ip{\psi(t)}{\bf x}\bra{\bf x}i[\hat{H},\hat{\bf x}]\ket{\psi(t)}}{\hbar \ip{\psi(t)}{\bf x}\ip{\bf x}{\psi(t)}}. \label{WVvel}
\eeq
For $\hat{H} = {\hat{\bf p}^2}/{2m} + V(\hat{\bf x})$, we have $i[\hat{H},\hat{\bf x}]/\hbar = \hat{\bf p}/m$, so \erf{WVvel} immediately reduces to the Bohmian velocity (\ref{BMvelocity}) for the standard probability current (\ref{madelung}).

The above \blu proof \blk is by no means limited to a single scalar particle. In a general configuration space, \erf{WVvel}  still holds, but now 
${\bf x}$ is  an infinite-dimensional vector. This vector  includes the 3-positions of all the particles (which may also have spin) and it also includes the values of all the quantized gauge fields (e.g.~the electromagnetic vector potential in the Coulomb gauge) at every point in space. Now it was shown in Ref.~\cite{GamWis04} that as long as $\hat{H}$ is  at 
most quadratic in the observables conjugate to $\hat{\bf x}$, 
\erf{WVvel} gives a velocity field for ${\bf x}$ that conserves the QM probabilities
by satisfying 
\beq \label{conservation}
\dbd{t} P({\bf x};t) + \sum_n \dbd{x_n}  [v_n({\bf x};t) P({\bf x};t)] = 0 .
\eeq
Note that $v_n$, the $n$th component of the velocity (that is, $\dot{x}_n$  in general BM),  
is given explicitly by
\beq \label{WVvel2}
{v}_n({\bf x};t) = {\rm Re}\frac{\ip{\Psi(t)}{\bf x}\bra{\bf x}i[\hat{H},\hat{x}_n]\ket{\Psi(t)}}{\hbar \ip{\Psi(t)}{\bf x}\ip{\bf x}{\Psi(t)}}.
\eeq
Here I use the capital $\ket{\Psi}$  to emphasize its universality. 

\section{Determinism: A Necessary Assumption}
In the above method for measuring ${\bf v}({\bf x};t)$, any individual weak measurement of velocity 
is very noisy. Thus strictly the weak value ${\bf v}({\bf x};t)$ should be interpreted (by a  naive  experimentalist, with no knowledge of QM or BM) only as the {\em mean} velocity in configuration space --- this noise could be masking variations in the velocity between individual systems that have the same Bohmian configuration ${\bf x}$ at time $t$. There are in fact other interpretations \cite{Nel66} in which ${\bf x}$ is the HV, but in which the motion of ${\bf x}$ is stochastic, and ${\bf v}({\bf x};t)$ {\em is} only the mean velocity. Thus to \blu obtain \blk BM from the measured ${\bf v}({\bf x};t)$ it is necessary to make the assumption of {\em determinism}. 
Given that assumption, the naive experimentalist  with a large enough ensemble could reconstruct the possible paths of Bohmian particles directly from experimental data.  For example, the set of Bohmian paths in the twin-slit experiment first calculated in Ref.~\cite{PhiDewHil79} are naively observable in this way. 

It must be emphasized that the technique of measuring weak values does not allow an experimenter  (naive or otherwise) to follow the path of an individual particle. That would violate the predictions of QM by, for example, enabling superluminal signalling \cite{Dic98}. Rather, as just stated, the above technique allows a naive experimenter to reconstruct all possible Bohmian paths from the measured velocity field ${\bf v}({\bf x};t)$ over all configuration space and over the time interval of interest, using a large ensemble of identically prepared quantum systems.  As stated in the Introduction, BM is an interpretation of QM, not a different theory, so {\em no} experiment can  distinguish BM from Orthodox QM.  What this experiment would establish is that the most natural operational definition of the configuration velocity (\ref{opdefvel}), with the assumption of determinism, \blu gives rise to paths in configuration space that match those of BM. \blk

\section{Position as the HV: An Unnecessary Assumption} \label{sec:posn}

There is an implicit assumption in the preceding paragraph, namely that the HV is the configuration {\bf x}, rather than some other physical quantity. 
At first sight it would seem that this is a necessary assumption for \blu motivating \blk BM by the above method, since any set of commuting observables can be chosen to give the HV \cite{Sud86,Bub97,GamWis04}. To obtain a deterministic theory (as we are assuming) it is necessary to restrict to observables with a continuous spectrum \cite{Bub97}  (p. 136). But this still allows for Bohmian-style deterministic trajectories for the momentum ${\bf p}$ as a HV \cite{BroHil01}.   Indeed, it allows for such trajectories for any linear symplectic transformation (LST) of the configuration ${\bf x}$. \footnote{ A LST of phase space $({\bf x},{\bf p}) \to ({\bf x}',{\bf p}')$ is defined by the equations ${\bf x}' = A{\bf x} + B{\bf p}$ and  ${\bf p}'=C{\bf x}+D{\bf p}$, where $AB^{\rm T}=BA^{\rm T}$, $CD^{\rm T}=DC^{\rm T}$ and $AD^{\rm T}-BC^{\rm T} = I$ \cite{Bro04}.} 

However, if we assume that the velocity  is given by the most natural operational definition,  \erf{WVvel2}, then we can rule out replacing $\hat{\bf x}$ by $\hat{\bf p}$. The reason is that $\hat{H}$ is {\em not} at most quadratic in $\hat{\bf x}$, the variables conjugate to $\hat{\bf p}$. Hence ${\bf v}({\bf p};t)$ derived from \erf{WVvel2} (with ${\bf x}$s changed to ${\bf p}$s) is incompatible with the evolution of $P({\bf p};t)$ in QM \cite{GamWis04}. For example, the Coulomb potential between particles is not quadratic in the particle separation. Similarly, there are in general non-quadratic interaction terms appearing in the Hamiltonian for gauge fields, such as those mediating the Strong Force \cite{Ryd85}. 

These facts  have a remarkable implication: It is in general not necessary to choose ${\bf x}$ as the HV \blu if one motivates BM by the above route. \blk Rather, the {\em asymmetry} between the configuration ${\bf x}$ and the conjugate momenta ${\bf p}$ in physical Hamiltonians makes ${\bf x}$ the only choice if the HV is assumed to evolve deterministically with a velocity which is  naively observable as a weak value. Specifically, this assumption rules out any LST ${\bf x}'$ that involves a component $p_j$ of the physical momentum for which the physical potential $V$ is not at-most-quadratic in the conjugate position $x_j$.  Of course for particular systems, such as harmonic oscillators in which ${\bf x}$ and ${\bf p}$ both appear quadratically, this argument cannot work. But when it is applied  to the general  sorts of Hamiltonians that we find in Nature, it singles out the configuration ${\bf x}$ as the choice of HV, and BM as its dynamics. 

It follows that, as long as one accepts the  natural operational definition of the velocity of the HV given in Sec.~\ref{Sec:WVPC}, the necessity of choosing ${\bf x}$ as the HV becomes a naively determinable fact. (This is assuming, as always, that   QM is correct in its experimental predictions.) That is, the naive experimentalist, with no knowledge of QM,  could verify  experimentally that the velocity of ${\bf p}$ in momentum space, defined as 
\beq
v({\bf p};t) = {\lim_{\tau\to 0} \,\tau^{-1}\, {\rm E}[{\bf p}_{\rm strong}(t+\tau) - {\bf p}_{\rm weak}(t)|{\bf p}_{\rm strong}(t+\tau) = {\bf p}]}, 
\eeq
is in general incompatible with the observed evolution of the momentum distribution $P({\bf p};t)$.  Similarly, any LST ${\bf x}'$  would also be ruled out as the HV --- only the physical configuration ${\bf x}$  would be found to give consistent evolution. 
%

\section{\blu Subjective Probabilities in BM} 

It is a simple consequence of the Bohmian dynamical law (\ref{WVvel2}) that the probability density $P({\bf x};t)$ for the HV ${\bf x}$ agrees with that of standard QM for all times  if it is assumed to hold at some initial time $t_0$.  While this assumption allows BM to replicate all the predictions of standard QM \cite{Boh52,BohHil93,Hol93,CusFinGol96,Dur01}, it prompts the question: why should the initial configuration have a probability density equal to $\ip{\Psi(t_0)}{\bf x}\ip{\bf x}{\Psi(t_0)}$? That is, why should $\ket{\Psi(t)}$  have a dual role, both guiding the configuration and determining its probability density? This question has been addressed  in the above references (see also Refs.~\cite{Val91,DurGolZan92}), but not, to my mind, answered completely satisfactorily. \blu It relates to the even deeper question, which also arises in other physical theories such as statistical mechanics, namely \blk what do we {\em mean} by a probability density $P({\bf x};t)$?

I believe the most fruitful viewpoint to adopt, \blu in BM as elsewhere, \blk is the subjective or 
Bayesian interpretation of probabilities. This approach is perhaps best summed up 
by the slogan  ``probability is not real'' \cite{deF74}. In the current context it means that the probability distribution $P({\bf x};t_0)$ for ${\bf x}$ at time $t_0$, known as the prior probability distribution (or prior for short),  is nothing but an expression of an observer's beliefs about ${\bf x}$, and observers may differ in their beliefs. \blu This contrasts with most treatments of probability in BM (see, for example, Valentini~\cite{Val91}), which make no reference to an observer's beliefs. The present article is not the place to extol all the merits of the Bayesian approach, but I hope at least to convince the reader that it is a {\em consistent} approach to adopt in the Bohmian context.\blk 


\blu At first sight there would seem to be a particular problem in adopting the Bayesian interpretation of probability in BM: how could such subjective probabilities give rise to the  probabilities  in standard QM, which are objective (at least when observers agree on the quantum state)? \blk However, there is a principle of Bayesian probability theory that, if followed,  suggests that all observers should agree that the prior for the Bohmian configuration is 
\beq \label{correctprior}
P_{\rm prior}({\bf x};t_0) = \ip{\Psi(t_0)}{\bf x}\ip{\bf x}{\Psi(t_0)}.
\eeq
I refer to the principle of indifference, so-called by Jaynes \cite{Jay03}: 
if the statement of a statistical problem is invariant under some transformation, then the prior 
distribution one chooses should respect this indifference.\footnote{\blu This principle of 
indifference is a more rigorous formulation of Laplace's principle of insufficient reason, and is a special case of the maximum entropy principle, also advocated by Jaynes \cite{Jay03}. The last principle is not universally accepted even among Bayesians \cite{Uff95}, but the principle of indifference as stated above is less controversial, and is all that is needed for the present argument.\blk} 
This can be applied as follows. We assume only the dynamics of BM:  the guidance equation (\ref{WVvel2}), and the time-dependent guiding function:  
\beq
\ket{\Psi(t)} = e^{-i\hat{H}(t-t_{0})/\hbar}\ket{\Psi(t_{0})}.
\eeq
Now there is no particular significance to the time $t_0$. Therefore the prior should be covariant with respect to translation in time. That is, 
\beq
\dbd{t}P_{\rm prior}({\bf x};t) = \sum_n \dbd{x_n}[ P_{\rm prior}({\bf x};t)\dot{x}_n({\bf x};t)].
\eeq
Given the equation for $\dot{\bf x}$ (\ref{WVvel2}) in terms of the guiding function 
$\ket{\Psi(t)}$, we have already shown that a prior distribution that satisfies this covariance equation is that of \erf{correctprior}. 

The information to which the problem is {\em not} indifferent is the guiding function and the guidance equation. Choosing the prior to be some other function of this information, such as 
$\ip{\Psi(t_0)}{\bf x}\ip{\bf x}{\Psi(t_0)}^{p}$ for $p\neq 1$, will not give a prior that is 
covariant with respect to time translations. This special covariance property of $P({\bf x};t)$ was emphasized some time ago by D\"urr, Goldstein and Zangh\`i \cite{DurGolZan92,DurGolZan96} (they call it equivariance of the equilibrium distribution).  No other covariant prior constructed solely from the guiding function and the guidance equation \blu is known  for a general Hamiltonian, so it is reasonable to suppose that $P({\bf x};t)$ is the unique  ``correct'' prior. \blk

Even if an observer were to ignore Jayne's principle of indifference, and instead chose some other prior, then for $t_0$ in the sufficiently distant past, it can be argued that this would not matter, as long as the prior was not pathologically different from \erf{correctprior}. The argument put by Valentini \cite{Val91} is along these lines: BM for realistic physical systems is chaotic, and there are analogs of the statistical mechanical $H$-theorems \cite{Tol79} showing that an arbitrary $P({\bf x},t)$ will, under the guidance equation, tend towards the equilibrium distribution $\ip{\Psi(t)}{\bf x}\ip{\bf x}{\Psi(t)}$.  This argument relies upon coarse-graining of the distribution. \blu This is easily justified for the case of subjective probability: \blk a real observer cannot  compute the trajectories for a chaotic system with arbitrary accuracy for any length of time. The reason is of course that any errors will grow exponentially, so that the observer's knowledge will invariably become coarser with time.

It \blu must be pointed out that D\"urr, Goldstein and Zangh\`i~\cite{DurGolZan92} do discuss subjective probability in the Bohmian context. However,  they do not favour this interpretation, saying merely that ``as a {\em consequence} [their emphasis] of our analysis, the reader, if he so wishes, can safely \ldots regard $P({\bf x};t_0)$ as \ldots a subjective probability for the initial configuration''.  The analysis to which they refer consists of showing that the Bohmian prediction for the frequency of measurement outcomes on a large ensemble of quantum systems with (macroscopically) identical preparation and measurement procedures {\em typically} agrees with the standard QM probabilities; ``Typicality is to be here understood in the sense of quantum equilibrium,'' \cite{DurGolZan92} that is, the distribution of \erf{correctprior}. This agreement with standard QM is presumably what they mean by saying that somebody with Bayesian inclinations can choose this prior  ``safely''. 


Regardless of whether one justifies $P({\bf x};t_0)$ as a measure of typicality by the empirical validity of QM, or by the arguments I have given above, \blk the analysis of D\"urr, Goldstein and Zangh\`i still holds. This analysis shows that if an observer $o$ had control over a particular quantum system $s$, to which she assigned (in agreement with other observers) the state $\ket{\psi_{s}}$, then she should (in agreement with other observers) assign the configuration for that system ${\bf x}_s$ the equilibrium distribution  $P_\psi({\bf x}_{s}) = \ip{\psi}{{\bf x}_{s}} \ip{{\bf x}_{s}}{\psi}$. Note that here $\ket{\psi_s}$ is a quantum state with the usual meaning in Orthodox QM: a means of predicting the outcomes of future measurements on that system. Because BM reproduces all the predictions of Orthodox QM, quantum states in this sense emerge from BM too. The differences in BM are that: first,  the observer is unambiguously defined, being made of particles and fields with a definite configuration ${\bf x}_o$; and second, $\ket{\psi_s}$ also serves to guide ${\bf x}_s$.
For example, if the universe comprised only $o$ and $s$, and $o$ could assign a pure state to $s$, then that state would be
\beq \label{relativestate}
\ket{\psi_s}\bra{\psi_s} \propto {\rm Tr}_o[ \ip{{\bf x}_o}{\Psi}\ip{\Psi}{{\bf x}_o}].
\eeq 
 \blu (Note that the partial trace here is necessary because the observer $o$ may have subsystems, 
such as spin, not spanned by $\cu{\ket{{\bf x}_o}}$.) 
The  observer can update her distribution for ${\bf x}_s$ from $P_\psi({\bf x}_{s})$ to something narrower by obtaining more information about the system. However, \blk this requires interacting with the system, and will change $\ket{\psi_s}$ proportionately, just as in Orthodox QM. 
 
Note that in \erf{relativestate} it is as if the observer knows her own configuration ${\bf x}_0$. Such a degree of self-knowledge is not realistic. \blu  Nor is it required  for  \erf{relativestate} to hold; if $\ket{\psi_s}$ is the known (by $o$) state of $s$, then this knowledge will be encoded in just a few macroscopic degrees of freedom of $o$. \blk  Nevertheless, because the observer is part of the universe  in BM,  her knowledge of ${\bf x}$ is certainly {\em not} limited to the equilibrium distribution for 
the universe as a whole. That is,  
\beq \label{notprior}
P({\bf x};t) \neq \ip{\Psi(t)}{{\bf x}} \ip{{\bf x}}{\Psi(t)},  \eeq
where ${\bf x}$ incorporates ${\bf x}_o$.\footnote{This point is not new, but has not, in my view, been sufficiently emphasized. For example it is made in Ref.~\cite{DurGolZan92}, but only as a ``random point'' in the Appendix.} The right-hand-side is only a {\em prior} distribution, which would be held only by a totally innocent observer. The left-hand-side is the {\em posterior} distribution that any real observer will have, conditioned upon her observations. As soon as an innocent observer were to open her eyes she would collapse her state of belief about ${\bf x}$  from $P_{\rm prior}({\bf x};t)$ to a much sharper $P({\bf x};t)$, by observing  the location of objects (from the pointer on a meter to the stars in the sky) relative to her. Note that this ``collapse'' is completely classical: it is just Bayesian updating of her beliefs about the positions of macroscopic objects. The guiding function $\ket{\Psi(t)}$ of course does not collapse\footnote{This Bayesian updating by an observer in BM is thus similar to the pruning of other branches by an observer in each branch of Everett's universal wavefunction \cite{Eve57}. The difference is that in BM there is a  unique real branch singled out by ${\bf x}$, and probabilities can be interpreted in the usual way.}. In BM one should not think of $\ket{\Psi(t)}$ as a quantum state in the usual sense but rather, as I have called it, a guiding function, \blu the essential constituent \blk of the law of motion (\ref{WVvel2}).  

\section{Discussion}

The crucial assumptions behind my claim that BM is preferred over other HV interpretations on physical grounds are that the HV dynamics are deterministic, and that the velocity-field of the HV should be naively observable, which means defining it operationally as a {\em weak value}. I showed in Sec. 4 that these assumptions are compatible with QM only if one chooses the physical configuration ${\bf x}$ as the HV. Hence  these assumptions lead uniquely to BM, as I showed in Sec. 2. In this section I discuss the relation between this route to BM, and previous work.   

It was pointed out by Holland \cite{Hol99} that,  if one takes the position $\vec{x}$ of a single electron obeying Dirac's equation as one's HV, the assumption  of relativistic covariance fixes a unique expression for the velocity, $\vec{v}={\vec j}/\rho$, where $({\vec j},c\rho)$ is the usual Dirac 4-current.  It is easy to verify that Holland's expression is reproduced by the general expression for the velocity as a weak value (\ref{WVvel2}), using the Dirac Hamiltonian. \blu 
It might seem that Holland's justification for his unique HV interpretation is simpler and more compelling than that which I have offered. \blk However, Holland's argument does not \blu justify $\vec x$ as the preferred HV, but rather simply assumes it. \blk Also, while Holland's argument provides a unique velocity for Dirac particles, it cannot be applied to obtain unique velocities for the values of a gauge field at all points in space. \footnote{ Note that it is not possible to define a particle-velocity from a relativistic equation for a bosonic field, analogous to the  particle-velocity for the single-particle Dirac wavefunction. That is because the time-component of the  4-current constructed from the energy-momentum tensor of the field does not remain positive for interacting fields, as required if it is to represent a probability density \cite{Str04}. }  

It is interesting to note, as Holland does, that in the non-relativistic limit his expression for $\vec v$ does not correspond to the usual Bohmian velocity (as derived in the present paper), but has an extra spin-dependent term.  The present paper shows that this term does not arise 
 using the operational definition of the velocity presented in Sec.~\ref{Sec:WVPC} using the non-relativistic Hamiltonian. But on the other hand, if one applies this operational definition using the Dirac equation, and then takes the non-relativistic limit, one will obtain Holland's extra term. The different answers thus arise from taking the non-relativistic limit before or after applying the weak-value formula (\ref{opdefvel}). 

Which answer would be observed experimentally? My analysis in Sec.~\ref{Sec:WVPC} assumed already the non-relativistic Hamiltonian, which is valid only for time-scales much greater than the zitterbewegung period, $T_{\rm zit} = h/2mc^2$. Thus if the times taken for the weak or strong measurements, or the the separation $\tau$ between them, were much greater than $T_{\rm zit}$, the standard Bohmian velocity is what would be measured experimentally. Since $T_{\rm zit} \approx 10^{-22}$s  for an electron, this is the relevant regime for the foreseeable future. For Holland's extra term to be experimentally observed by the method of the present paper, all of the above time scales would have to be much less than $T_{\rm zit}$. In this limit the complementarity between energy and time implies that pair creation could not be avoided, so the attempt to measure the Bohmian velocity of a single electron would necessarily fail. 

A different means of  deriving a unique expression for ${\bf j}$, and hence ${\bf v}$, 
is the algebraic approach of Brown and Hiley \cite{BroHil01}. They claim that BM  ``arises directly from the non-commutative structure of quantum phase space.''  Their  
 method is not restricted to the configuration ${\bf x}$; as noted 
in Sec.~\ref{sec:posn}, they also construct trajectories for the momentum ${\bf p}$. 
However, in subsequent work \cite{Bro04}, Brown has identified a ``mechanical velocity'' 
in addition to the ``current velocity'' ${\bf v}$. He finds that, for systems with a higher-than-quadratic potential, these velocities disagree unless ${\bf x}$ is chosen as the HV. Thus he  concludes that ``the co-ordinate representation therefore appears special for the Bohmian interpretation.'' Note that the velocities defined by Brown and Hiley are not defined 
operationally, unlike the velocity I introduced in the present paper.

Finally, it should be noted that in Bohm's work \cite{Boh52,BohHil93}, the velocity ${\bf v}$ of ${\bf x}$ was not derived from the current ${\bf j}$. Rather, it was defined by developing an analogy between the \sch\ equation for $\psi({\bf x})$ and the Hamilton-Jacobi equation for $S({\bf x})$, plus a continuity equation for $P({\bf x})$. Bohm's approach also singles out the position as the preferred HV, because the Hamilton-Jacobi equation is derived ultimately from a Lagrangian, a function of ${\bf x}$ and $\dot{\bf x}$. However, Bohm's approach cannot be applied to deriving a velocity from the Dirac equation, because the Dirac Hamiltonian operator, being linear in $\hat{p}$, cannot be derived from a Lagrangian. By contrast, the weak-valued velocity (\ref{WVvel2}) works for any fundamental physical Hamiltonian, including Dirac's, as noted above.

\section{Conclusion}
In this article I have shown the following. (i) The  probability current in configuration space  has a natural operational definition using weak measurements. (ii) This operational definition agrees with the standard expression for the quantum probability current. (iii) As a consequence, the possible trajectories of the hidden configuration ${\bf x}$ in the Bohmian interpretation of QM can be determined by a naive experimentalist knowing only that this interpretation is deterministic.  (iv) Adopting this operational definition for the velocity of a hidden variable in general, the asymmetry between position and momentum in physical Hamiltonians singles out the former. That is, if the trajectories are to be compatible with the experimentally observable evolution of the probability distribution, the hidden variable must be the configuration ${\bf x}$ as in Bohmian mechanics. (v) Given the Bohmian guidance equation for ${\bf x}$, the usual quantum distribution for ${\bf x}$ can be \blu motivated in the context of Bayesian probability theory as the unique prior covariant under translation of the initial time. \blk

This work has not attempted to answer every question relating to Bohmian mechanics. However it  has made substantial  progress in connecting it to experiment, identifying its unique attributes among hidden variables interpretations, and justifying its foundations.

\section{Acknowledgments} I would like to acknowledge discussions with Rob Spekkens, Michael Hall, Mitchell Porter, Rick Leavens, and Basil Hiley. This work was supported by the Australian Research Council. 

\end{document}